\documentclass[aps,prb,floatfix,superscriptaddress]{revtex4-2}
\usepackage{graphicx,bm,hyperref,color,multirow,amsmath}
\newcommand{\vect}[1]{\boldsymbol{#1}}
\begin{document}

\title{Fe-site-resolved anisotropy energies in Nd\textsubscript{2}Fe\textsubscript{14}B for atomistic spin dynamics}
\author{Veronica T.\ C.\  Lai}
\affiliation{
Department of Materials, University of Oxford,
Parks Road,
Oxford OX1 3PH, United Kingdom}
\author{Christopher E.\ Patrick}
\email{c.patrick.1@warwick.ac.uk}
\affiliation{
Department of Materials, University of Oxford,
Parks Road,
Oxford OX1 3PH, United Kingdom}
\affiliation{
Department of Physics, University of Warwick, Gibbet Hill Road, 
Coventry CV4 7AL, United Kingdom}

\date{\today}

\begin{abstract}
Nd-Fe-B magnets are the most widely used high performance
magnets in the world today, and remain the subject of both
experimental and computational research aimed at understanding
and optimizing them.
Atomistic spin dynamics (ASD) is one technique which has been
used in recent years to provide insight
into magnetic properties relevant to coercivity, such as domain
wall width.
Although it is relatively clear how to model magnetocrystalline
anisotropy arising from rare-earth atoms in these simulations, 
the contribution from the transition metal Fe is less obvious,
due to the itinerant nature of the magnetism.
Here, we examine previous treatments of Fe anisotropy in ASD simulations
and identify a discrepancy with previously-published first-principles
studies.
We derive two models which correct this discrepancy, one based on single-ion theory and the other
on anisotropic exchange,
and test their performance by comparing to first-principles torque calculations
on Y$_2$Fe$_{14}$B.
The torque calculations show a contribution which cannot be explained
by the single-ion model but arises naturally from (antisymmetric) anisotropic
exchange.
We propose practical strategies to model Fe anisotropy
in future ASD simulations, including a simplified (mean-field) description
of anisotropic exchange, which may have applications beyond R$_2$Fe$_{14}$B
to the wider class of itinerant magnetic materials.
\end{abstract}

\maketitle

\section{Introduction}
\label{sec.intro}

Rare-earth permanent magnets, particularly those based
on Nd$_2$Fe$_{14}$B~\cite{Herbst1985,Sagawa1984} continue to 
dominate the global market due to their unmatched performance~\cite{Nakamura2017,Gutfleisch2011}.
Due to their widespread use, even incremental improvements
to their magnetic properties and/or reduction in their rare-earth content can
have significant impact, with the development of the grain-boundary
diffusion process being an important example~\cite{Kim2019}.
As such, they remain a topic of active research, both
experimentally and computationally.
Indeed, due to the inherent complexity of Nd-Fe-B magnets,
it is only relatively recently that computational methods
have advanced to the stage where one can carry out
realistic simulations, whether at the
first-principles (most commonly density-functional theory, DFT),
atomistic spin dynamics (ASD) or micromagnetics level.

A key computational challenge, particularly associated with understanding
the extrinsic processes of magnetization reversal and coercivity, is 
accurately describing magnetic domain walls, which due to the extremely
high magnetocrystalline anisotropy energy (MAE) are expected to be
only a few nm wide~\cite{Herbst1991}.
The fact that reverse domains are only separated by a few atomic layers
would seem to mandate an atomistic description of the magnetism,
as supplied by DFT or ASD, rather than a continuum micromagnetic description.
However, the high computational cost of DFT currently prohibits explicit
modeling of domains, leaving 
ASD to provide a unique opportunity to model domain walls
and coercivity mechanisms,
with a number of important and interesting 
studies having been carried out
in recent years~\cite{Toga2016,Nishino2024,
Nishino2022,Nishino2021,Nishino20212,Uysal2020,Nishino2017,
Gong2020,Gong20192,Gong2019}.

The dominant contribution to the anisotropic properties
of rare-earth magnets is from the lanthanide elements themselves, 
namely the crystal-field interaction of the asymmetric $4f$-electron
charge cloud with its surroundings~\cite{Kuzmin2008}.
This arguably has been well understood since the 
development of crystal field and Callen-Callen theories~\cite{Newman1989,Callen1960}.
However, it has also been argued that
the contribution from the Fe atoms is important~\cite{Bouaziz2023},
particularly at high temperature.
More generally, as research efforts continue in the field
of rare-earth-lean and rare-earth-free magnets, the accuracy
of ASD models of these systems will rely critically on a realistic 
description of transition metal anisotropy.
Previous ASD studies of $R_2$Fe$_{14}$B (where $R$ usually denotes
Nd)~\cite{Toga2016,Nishino2024,
Nishino2022,Nishino2021,Nishino20212,Uysal2020,Nishino2017,
Gong2020,Gong20192,Gong2019} 
have employed the following expression
for the anisotropy energy at the Fe sites, $E_{s,i}$,
\begin{equation}
    E_{s,i}(\vect{S_i}) = -D_iS_{iz}^2 
    \label{eq.Di}
\end{equation}
where the $D_i$ coefficients are described as anisotropy parameters, and
$S_{iz}$ denotes the $z$ ($c$-axis)-component of a unit vector 
$\vect{S_i}$ describing the orientation of the magnetic moment at
site $i$.

As discussed in Section~\ref{sec.theory},
critical comparison of Eq.~\ref{eq.Di} 
with previously-published first-principles studies
of Y$_2$Fe$_{14}$B shows that it
provides an incomplete 
description of the Fe anisotropy.
This has motivated us to carry out this present
study, where we derive models
of the Fe anisotropy in $R_2$Fe$_{14}$B
to be used in ASD simulations which are
consistent with the previously-published first-principles
studies.
We have tested our models by carrying out new
first-principles calculations of the atom-resolved torques
in Y$_2$Fe$_{14}$B.
These calculations reveal an additional torque contribution
which cannot be explained by a model based only on 
single-ion anisotropy.
However, the additional torque contribution does arise
naturally when the anisotropy of the exchange interaction
is taken into account.
Observing that parameterizing the full exchange tensor
may be prohibitively expensive for a complex crystal
structure like $R_2$Fe$_{14}$B, we propose a simplified
expression to describe the anisotropic exchange using
single-site-resolved quantities.

The rest of the manuscript is organized as follows.
Section~\ref{sec.theory} describes the Hamiltonian 
used in previous ASD simulations, with particular focus
on the description of Fe anisotropy and an
inconsistency with previous first-principles studies.
We present two alternative models which
address this issue.
In Sec.~\ref{sec.results} we evaluate the performance
of the models against new DFT calculations on
Y$_2$Fe$_{14}$B.
In Sec.~\ref{sec.discussion} we discuss our findings
and present practical strategies to carry out ASD simulations
in future with an improved description of Fe anisotropy.
We give our conclusions in Sec.~\ref{sec.conclusions}.

\section{Theory}
\label{sec.theory}

\subsection{The standard ASD Hamiltonian}

The most widely used approach to modeling the spin dynamics
of Nd\textsubscript{2}Fe\textsubscript{14}B and related
materials is to employ an ASD Hamiltonian of the following
form~\cite{Toga2016,Nishino2024,
Nishino2022,Nishino2021,Nishino20212,Uysal2020,Nishino2017,
Gong2020,Gong20192,Gong2019}:
\begin{eqnarray}
    \mathcal{H} &=& -\sum_{i} \sum_{j > i}J_{ij}\vect{S_i}\cdot\vect{S_j}
    +\sum_{i\in \mathrm{Fe}}E_{s,i} (\vect{S_i}) \nonumber \\
    && + \sum_{i\in R} E_{Rs,i}(\vect{S_i}) - \mu_0\sum_i{m_i}\vect{h}\cdot\vect{S_i}
    \label{eq.HASD}
\end{eqnarray}
with the four terms accounting respectively for the inter-spin
exchange interaction (assumed pairwise and isotropic), the anisotropy
of the Fe atoms, the anisotropy of the $R$ atoms, and the Zeeman interaction
with the external magnetic field $\vect{h}$.
The detailed form of $\mathcal{H}$ varies between different
implementations depending on, for instance, whether an additional
factor of 2 is included in the exchange 
parameter $J_{ij}$, or whether the $\vect{S_i}$
vectors represent both the magnitude and direction of the 
magnetic moments, or just the direction.
For definiteness, in Eq.~\ref{eq.HASD}
$m_i$ is the magnitude of the local magnetic moment
at site $i$, and $\vect{S_i}$ is a unit vector which gives its orientation.
The orientation can be described equivalently through the 
angles $(\theta_i,\phi_i)$ in the usual way,
\begin{equation*}
    \vect{S_i} = 
 \begin{pmatrix}
        S_{ix} \\
        S_{iy} \\
        S_{iz} 
    \end{pmatrix}
    =
    \begin{pmatrix}
        \sin\theta_i\cos\phi_i\\
        \sin\theta_i\sin\phi_i\\
        \cos\theta_i
    \end{pmatrix}.
\end{equation*}

The single-ion anisotropy of the $R$ atoms $E_{Rs,i}(\vect{S_i})$
is given by a relatively complicated expression involving Stevens
operators and crystal field coefficients.
It originates from an angular $(l,m)$ expansion of the 
crystal field potential at the $R$ site, and usually all terms
in the expansion which couple to the 4$f$ electrons 
($l=2,4,6$) are included~\cite{Kuzmin2008}.
The point symmetry of a given site $i$ determines which crystal
field coefficients can be nonzero, which sets the functional
form of $E_{Rs,i}$.
Since our focus in this work is on the Fe atoms, we will not discuss this 
term further, nor the Zeeman term whose functional form is well known.

\subsection{Single-ion Fe anisotropy in previous works}

As stated in Section~\ref{sec.intro}, previous works
have used Eq.~\ref{eq.Di} to model the anisotropy
energy at the Fe sites.
The cited studies~\cite{Toga2016,Nishino2024,
Nishino2022,Nishino2021,Nishino20212,Uysal2020,Nishino2017,
Gong2020,Gong20192,Gong2019} take the values of the $D_i$ parameters 
from the paper by Miura et al., Ref.~\cite{Miura2014}.
In that work, the MAE
of Y\textsubscript{2}Fe\textsubscript{14}B was calculated
using DFT in the 
projector-augmented wave (PAW) formalism.
A value of 0.622~meV/FU (formula unit) 
was obtained through the magnetic force theorem,
as the difference in single particle eigenvalues calculated
with all moments aligned to the [001] direction ($S_{iz}=1$)
and with them all aligned to the [100] direction ($S_{iz} =0$).
According to the ASD Hamiltonian and the assumed form
of the anisotropy energy (Eqs.~\ref{eq.Di}~and~\ref{eq.HASD}) 
and noting that $E_{Rs,i}$ is zero for $R$ = Y, we
see that the energy difference $E^{[100]} - E^{[001]} = \sum_i{D_i}$.
Crucially, Ref.~\cite{Miura2014} also provided an estimate 
of the site-resolved anisotropy energies, obtained by
weighting the eigenvalues by the PAW projections of their
respective eigenstates.
These values were reported in Fig.~1 of Ref.~\cite{Miura2014}
for the different Fe sublattices, and have been tabulated
in subsequent ASD works, e.g.\ Ref.~\cite{Toga2016}.
The 56 Fe atoms in the conventional unit cell are assigned
one of six $D_i$ values, depending on which of the six Wyckoff sublattices
($16k_1$, $16k_2$,
$8j_1$, $8j_2$, $4c$ and $4e$) they belong to.
These $D_i$ values are 0.55, 0.38, 1.07, 0.58,
-2.14 and -0.03~meV/atom, respectively~\cite{Toga2016}.

Two important considerations arise regarding the approach taken
in Refs~\cite{Toga2016,Nishino2024,
Nishino2022,Nishino2021,Nishino20212,Uysal2020,Nishino2017,
Gong2020,Gong20192,Gong2019}.
The first is that if the MAE is calculated as a sum
over these $D_i$ values, properly weighted according
to the relative number of Fe atoms in the sublattices,
a value of 4.85~meV/FU is obtained.
This is more than seven times
greater than the value of 0.622~meV/FU
reported from the same DFT calculations prior to 
carrying out the PAW decomposition.
The second consideration is based on Ref.~\cite{Torbatian2014},
which also calculated the MAE of Y\textsubscript{2}Fe\textsubscript{14}B
using DFT.
In this case, a value of 0.65~meV/FU was obtained,
agreeing well with Ref.~\cite{Miura2014}.
The site-resolved anisotropy energies were also estimated
in Ref.~\cite{Torbatian2014}, but crucially, the work shows how
not all atoms in the same sublattice make the same contribution
to the MAE.
Specifically, the $16k_1$, $16k_2$ and $4c$ sublattices 
are each further split into two groups.
If one sums up the estimated decompositions from Ref.~\cite{Torbatian2014},
a value of 0.95~meV/FU is obtained; still somewhat larger than
the reported total of 0.65~meV/FU, but far closer than the
discrepancy noted for Ref.~\cite{Miura2014}.

Subsequent correspondence~\cite{Miuraemail} with the authors of 
Ref.~\cite{Miura2014} confirmed that they also observed
the splitting of the $16k$ and $4c$ sublattices, but
were not able to show all the values in Fig.~1 of
their manuscript due to limited space.
Such splittings cannot be explained by
Eq.~\ref{eq.Di}, assuming a single $D_i$ value
per Wyckoff sublattice.
This observation motivated us
to derive an alternative description of the Fe
anisotropy which does account for the splitting
observed in the 
previously-reported first-principles 
calculations~\cite{Torbatian2014,Miuraemail}.
Here, we report two different models.
The first model is closely aligned to 
Eq.~\ref{eq.HASD},
assuming an isotropic exchange interaction and single-ion
anisotropy.
However, rather than using Eq.~\ref{eq.Di}, we 
carefully construct the functional forms of $E_{s,i}(\vect{S_i})$
which are consistent with the symmetry of each site $i$.
The resulting expressions naturally account for the observed splittings.
However, when testing these expressions against new first-principles
calculations of anisotropy via the torque method rather than total
energy differences, we observe additional torques which cannot
be explained by a single-ion anisotropy.
This leads us to propose a second model where Eq.~\ref{eq.HASD}
is modified to include anisotropy in the exchange interaction.
We now describe these two models.

\subsection{Model 1: Single-ion anisotropy}

Model 1 uses the same spin Hamiltonian as shown in Eq.~\ref{eq.HASD}.
However, we generalize Eq.~\ref{eq.Di} to allow
for a more complicated angular dependence:
\begin{eqnarray}
E_{s,i}(\vect{S_i}) &=& \mathcal{A}_i \cos^2 \theta_i 
                   +\mathcal{B}_i \sin(2\theta_i) \cos\phi_i  \nonumber \\
&&                 +\mathcal{C}_i \sin(2\theta_i) \sin\phi_i 
                   +\mathcal{E}_i \sin^2\theta_i \cos(2\phi_i)  \nonumber \\
&&                 +\mathcal{F}_i \sin^2\theta_i \sin(2\phi_i)
\label{eq.angle2}
\end{eqnarray}
Equation~\ref{eq.angle2} is an expansion over the $l=2$ spherical
harmonics.
Neglecting all terms except the first and setting $\mathcal{A}_i = -D_i$
retrieves Eq.~\ref{eq.Di}.
The full expansion of Eq.~\ref{eq.angle2} 
in principle requires five coefficients per Fe site
(i.e.\ 280 in the $R_8$Fe$_{56}$B$_4$ unit cell).
However, making the assumption of single-ion anisotropy, the point
symmetry of site $i$ restricts the allowed form of $E_{s,i}$
in two ways.
First, if the crystal environment of atom $i$ is invariant under 
some symmetry operation, then $E_{s,i}$ must be invariant under
the same symmetry operation.
This leads to some of the coefficients in Eq.~\ref{eq.angle2}
vanishing at given sites.
Second, if a symmetry operation transforms 
the environment of site $i$
into that of another site $j$ (which will be a member of the same
Wyckoff sublattice), then  $E_{s,j}$ will be related to  $E_{s,i}$
through the same transformation.
As we shall show, this leads to there being a single set of 
coefficients per sublattice, with differences between $E_{s,i}$
within the sublattice occurring through changes in the signs
and ordering of the coefficients.

As an example, we take the 4$e$ sublattice of 
$R_2$Fe$_{14}$B, which has atoms
located at fractional positions $(0,0,z)$, $(0.5,0.5,0.5-z)$,
$(0,0,-z)$ and $(0.5,0.5,0.5+z)$, where 
$z= 0.1144$~\cite{Isnard1995}.
Applying the above symmetry arguments to the atom at $(0,0,z)$,
we obtain the following expression for its single-ion anisotropy:
\begin{eqnarray}
E_{s,4e,1}(\theta,\phi) =
\kappa_{20}\cos^2 \theta + \kappa_{22}\sin^2\theta\sin2\phi \nonumber
\end{eqnarray}
where $\kappa_{20}$ and $\kappa_{22}$ are constants.
The atom at $(0,0,-z)$ has an identical expression, $E_{s,4e,1}=E_{s,4e,3}$,
whilst the remaining two atoms' expressions are given by
\begin{eqnarray}
E_{s,4e,2}(\theta,\phi) = E_{s,4e,4}(\theta,\phi) = 
\kappa_{20}\cos^2 \theta - \kappa_{22}\sin^2\theta\sin2\phi \nonumber 
\end{eqnarray}
Importantly, summing over 
all sites in the sublattice retrieves the conventional uniaxial
angular dependence of Eq.~\ref{eq.Di}, proportional to $\cos^2\theta$ and independent of $\phi$.

\begin{table*}
\begin{tabular}{|c|c|c|c|c|c|c|c|c|c|c|}
\hline
Wyckoff Label&Index&$x$&$y$&$z$ & $\mathcal{A}_i$ & $\mathcal{B}_i$ & $\mathcal{C}_i$ & $\mathcal{E}_i$ &$\mathcal{F}_i$ & $\mathbf{D_i}$ \\
&&&&&$(-3\Delta\mathcal{J}_{i\parallel}/4)$&$(-\mathcal{J}_{ixz}/2)$&$(-\mathcal{J}_{iyz}/2)$&$( -\Delta\mathcal{J}_{i\perp}/2)$&$(-\mathcal{J}_{ixy}/2)$&\\
\hline
\multirow{ 8}{*}{$16k$}
&1&$x$     &$y$    &$z$    & $+\kappa_{20}$ & $+\kappa_{21c}$ & $+\kappa_{21s}$  & $+\kappa_{22c}$& $+\kappa_{22s}$&$(D_x,D_y,D_z)$ \\
&2&$0.5-x$ &$0.5+y$&$0.5-z$& $+\kappa_{20}$ & $+\kappa_{21c}$ & $-\kappa_{21s}$  & $+\kappa_{22c}$& $-\kappa_{22s}$&$(-D_x,D_y,-D_z)$ \\
&3&$0.5+x$ &$0.5-y$&$0.5-z$& $+\kappa_{20}$ & $-\kappa_{21c}$ & $+\kappa_{21s}$  & $+\kappa_{22c}$& $-\kappa_{22s}$&$(D_x,-D_y,-D_z)$ \\
&4&$-x$    &$-y$   &$z$    & $+\kappa_{20}$ & $-\kappa_{21c}$ & $-\kappa_{21s}$  & $+\kappa_{22c}$& $+\kappa_{22s}$&$(-D_x,-D_y,D_z)$ \\
&5&$y$     &$x$    &$-z$   & $+\kappa_{20}$ & $-\kappa_{21s}$ & $-\kappa_{21c}$  & $-\kappa_{22c}$& $+\kappa_{22s}$&$(D_y,D_x,-D_z)$ \\
&6&$0.5+y$ &$0.5-x$&$0.5+z$& $+\kappa_{20}$ & $+\kappa_{21s}$ & $-\kappa_{21c}$  & $-\kappa_{22c}$& $-\kappa_{22s}$&$(D_y,-D_x,D_z)$ \\
&7&$0.5-y$ &$0.5+x$&$0.5+z$& $+\kappa_{20}$ & $-\kappa_{21s}$ & $+\kappa_{21c}$  & $-\kappa_{22c}$& $-\kappa_{22s}$&$(-D_y,D_x,D_z)$ \\
&8&$-y$    &$-x   $&$-z$   & $+\kappa_{20}$ & $+\kappa_{21s}$ & $+\kappa_{21c}$  & $-\kappa_{22c}$& $+\kappa_{22s}$&$(-D_y,-D_x,-D_z)$ \\
\hline
\multirow{ 8}{*}{$8j$}
	&1&$x$     &$x$    &$z$    & $+\kappa_{20}$ & $+\kappa_{21}$ & $+\kappa_{21}$  &---& $+\kappa_{22}$& $(-D_{xy}, D_{xy},0)$ \\
	&2&$0.5-x$ &$0.5+x$&$0.5-z$& $+\kappa_{20}$ & $+\kappa_{21}$ & $-\kappa_{21}$  &---& $-\kappa_{22}$& $( D_{xy}, D_{xy},0)$ \\
	&3&$0.5+x$ &$0.5-x$&$0.5-z$& $+\kappa_{20}$ & $-\kappa_{21}$ & $+\kappa_{21}$  &---& $-\kappa_{22}$& $(-D_{xy},-D_{xy},0)$ \\
	&4&$-x$    &$-x$   &$z$    & $+\kappa_{20}$ & $-\kappa_{21}$ & $-\kappa_{21}$  &---& $+\kappa_{22}$& $( D_{xy},-D_{xy},0)$ \\
	&5&$x$     &$x$    &$-z$   & $+\kappa_{20}$ & $-\kappa_{21}$ & $-\kappa_{21}$  &---& $+\kappa_{22}$& $( D_{xy},-D_{xy},0)$ \\
	&6&$0.5+x$ &$0.5-x$&$0.5+z$& $+\kappa_{20}$ & $+\kappa_{21}$ & $-\kappa_{21}$  &---& $-\kappa_{22}$& $( D_{xy}, D_{xy},0)$ \\
	&7&$0.5-x$ &$0.5+x$&$0.5+z$& $+\kappa_{20}$ & $-\kappa_{21}$ & $+\kappa_{21}$  &---& $-\kappa_{22}$& $(-D_{xy},-D_{xy},0)$ \\
	&8&$-x   $ &$-x$   &$-z$   & $+\kappa_{20}$ & $+\kappa_{21}$ & $+\kappa_{21}$  &---& $+\kappa_{22}$& $(-D_{xy}, D_{xy},0)$ \\
\hline
\multirow{ 4}{*}{$4c$}
&1&$0$     &$0.5$  &$0$    & $+\kappa_{20}$ & --- &  ---  & $+\kappa_{22c}$&$+\kappa_{22s}$& $(0,0,D_z)$ \\
&2&$0.5$   &$0$    &$0.5$  & $+\kappa_{20}$ & --- &  ---  & $+\kappa_{22c}$&$-\kappa_{22s}$& $(0,0,-D_z)$ \\
&3&$0.5$   &$0$    &$0$    & $+\kappa_{20}$ & --- &  ---  & $-\kappa_{22c}$&$+\kappa_{22s}$& $(0,0,-D_z)$ \\
&4&$0$     &$0.5$  &$0.5$  & $+\kappa_{20}$ & --- &  ---  & $-\kappa_{22c}$&$-\kappa_{22s}$& $(0,0,D_z)$ \\
\hline
\multirow{ 4}{*}{$4e$}
&1&$0$     &$0$  &$z$    & $+\kappa_{20}$ & --- &  ---  & --- &$+\kappa_{22}$& --- \\
&2&$0.5$     &$0.5$  &$0.5-z$    & $+\kappa_{20}$ & --- &  ---  & --- &$-\kappa_{22}$& --- \\
&3&$0$     &$0$  &$-z$    & $+\kappa_{20}$ & --- &  ---  & --- &$+\kappa_{22}$& --- \\
&4&$0.5$     &$0.5$  &$0.5+z$    & $+\kappa_{20}$ & --- &  ---  & --- &$-\kappa_{22}$& --- \\
\hline
\end{tabular}
\caption{Derived relationships between anisotropy constants for the Fe sublattices in 
$R$\textsubscript{2}Fe\textsubscript{14}B.  For the $16k$ sublattice, note that atoms
9--16 are inversion partners of atoms 1--8 and have identical anisotropy constants.
For instance atom 9 in the $16k$ sublattice 
has coordinates $(-x,-y,-z)$, and has identical anisotropy constants
and $\mathbf{D_i}$-vector to atom 1. The ``---'' symbol implies a nonzero
value is disallowed by symmetry.  The coefficient values obtained by fitting
are presented in Table~\ref{tab.values}.
\label{tab.expressions}}
\end{table*}

Applying the same procedure to all of the Fe sites in 
$R_2$Fe$_{14}$B yields the following equations
for a particular site in each sublattice:
\begin{eqnarray}
E_{s,16k,1}(\theta,\phi) &=& 
\kappa_{20}\cos^2 \theta
+\kappa_{21c}\sin2 \theta\cos\phi \nonumber \\
&&+\kappa_{21s}\sin2 \theta\sin\phi
+ \kappa_{22c}\sin^2\theta\cos2\phi \nonumber \\
&&+ \kappa_{22s}\sin^2\theta\sin2\phi \nonumber \\
E_{s,8j,1}(\theta,\phi) &=& 
\kappa_{20}\cos^2 \theta
+\kappa_{21}\sin2 \theta (\cos\phi + \sin\phi) \nonumber \\
&&
+ \kappa_{22}\sin^2\theta\cos2\phi \nonumber \\
E_{s,4c,1}(\theta,\phi) &=& 
\kappa_{20}\cos^2 \theta 
+ \kappa_{22c}\sin^2\theta\cos2\phi \nonumber \\
&&+ \kappa_{22s}\sin^2\theta\sin2\phi \label{eq.fit_alltorques} 
\end{eqnarray}
Table~\ref{tab.expressions} provides a convenient representation
of the expressions for all sites in the sublattices.
The values of the coefficients in Eq.~\ref{eq.angle2}
are obtained by reading the row for the relevant atom
(the $\mathbf{D_i}$-vector column applies to Model 2 and
is explained in Sec.~\ref{sec.model2}).
We see that the low symmetry of the $16k$ sites 
lead to nonzero values for all five coefficients in 
Eq.~\ref{eq.angle2}, whilst for the $4e$ example
above, only two were nonzero.
Each sublattice ($16k_1$, $8j_2$, etc.) 
has its own set of $\kappa_{lm}$, and we provide
values for these based on DFT torque calculations in
Sec.~\ref{sec.results}, Table~\ref{tab.values}.
Altogether, the symmetry arguments reduce the number
of coefficients required to describe all of the Fe atoms from 
280 to 21.
Furthermore, as demonstrated 
for the $4e$ sublattice, averaging over all
atoms in a sublattice retrieves the simpler angular
dependence of Eq.~\ref{eq.Di} with $D_i = -\kappa_{20}$.

\subsection{Model 2: Anisotropic exchange}
\label{sec.model2}
Our second model is based on the following premise: the theory
of single-ion anisotropy arises naturally when magnetic moments
are associated with localized electrons and leads to predicted
behavior e.g.\ of the anisotropy constants with temperature,
which can be verified experimentally~\cite{Callen1960}.
However, the Fe magnetism is itinerant in nature, and its
temperature dependence (as measured for Y\textsubscript{2}Fe\textsubscript{14}B)
is known not to obey single-ion behavior~\cite{Hirosawa1986,Bouaziz2023}.
Therefore, we consider modifying the ASD Hamiltonian of
Eq.~\ref{eq.HASD} so that it leads to magnetic anisotropy
even in the absence of a single-ion contribution.
We do this by replacing the exchange term in Eq.~\ref{eq.HASD}
with a tensor form which is capable of describing an anisotropic
exchange interaction, i.e.\
\begin{equation*}
      -\sum_{i}\sum_{j>i} \vect{S_i} \vect{J_{ij}} \vect{S_j}.
\end{equation*}
Without loss of generality, we can separate out the isotropic 
part of $\vect{J_{ij}}$ to recover the original exchange 
term, plus a correction involving traceless tensors
$\vect{\mathcal{J}_{ij}}$,
\begin{eqnarray}
    -\sum_{i}\sum_{j>i} \vect{S_i} \vect{J_{ij}} \vect{S_j} &=&
    -\sum_{i}\sum_{j>i} J_{ij}\vect{S_i}\cdot \vect{S_j} \nonumber \\
     &&-\sum_{i}\sum_{j>i} \vect{S_i} \vect{\mathcal{J}_{ij}} \vect{S_j}.
    \label{eq.anisofull}
\end{eqnarray}

Determining all of the components of all of 
the $\vect{\mathcal{J}_{ij}}$ tensors would be a formidable task
for a material as complex as $R_2$Fe$_{14}$B.
However, we can rewrite Eq.~\ref{eq.anisofull} as follows, anticipating
its application to a ferromagnetic reference state.
We decompose the spin 
vectors into some global vector $\vect{M}$ plus a fluctuation
term, $\vect{S_{i}}=\vect{M + \vect{\delta S_i}} $.
The resulting modified exchange term can be written as:
\begin{eqnarray}
    -\sum_{i}\sum_{j>i}  \vect{S_i} \vect{J_{ij}} \vect{S_j} &=&
     -\sum_{i}\sum_{j>i}  J_{ij}\vect{S_i}\cdot \vect{S_j} \nonumber \\
     &&-\left(\sum_i \vect{S_i}\vect{\mathcal{J}_i}\vect{M}
     -\frac{1}{2} \sum_i\vect{M}\vect{\mathcal{J}_i}\vect{M} \right. \nonumber \\
     &&\left. + \frac{1}{2} \sum_i \sum_{j\neq i}
     \vect{\delta S_i}\vect{\mathcal{J}_{ij}}\vect{\delta S_j}
     \right)
     \label{eq.anisosimp}
\end{eqnarray}
The new quantities in Eq.~\ref{eq.anisosimp}
are the single-site exchange tensors $\vect{\mathcal{J}_i}$,
simply obtained as
\begin{equation*}
    \vect{\mathcal{J}_i} = \sum_{j\neq i}\vect{\mathcal{J}_{ij}}
\end{equation*}
A physical interpretation of $\vect{\mathcal{J}_i}$
is obtained from Eq.~\ref{eq.anisofull}, where
it appears in the anisotropic exchange contribution
from the $i$\textsuperscript{th} spin when all 
other spins are pointing along a common direction $\vect{M}$,
\begin{equation*}
    -\vect{S_i} \sum_{j\neq i} \vect{\mathcal{J}_{ij}} \vect{M}
= -\vect{S_i} \vect{\mathcal{J}_{i}}\vect{M}
\end{equation*} 

It is convenient to decompose $\vect{\mathcal{J}_i}$ into
a symmetric and antisymmetric part, 
$\vect{\mathcal{J}_i} = \vect{\mathcal{J}^s_i} + \vect{\mathcal{J}^a_i} $,
with
\begin{equation}
    \vect{\mathcal{J}^s_i}  = \begin{pmatrix}
     \Delta \mathcal{J}_{i\perp} - \frac{\Delta \mathcal{J}_{i\parallel}}{2}& \mathcal{J}_{ixy} & \mathcal{J}_{ixz} \\
\mathcal{J}_{ixy} &  -\Delta \mathcal{J}_{i\perp} - \frac{\Delta \mathcal{J}_{i\parallel}}{2} & \mathcal{J}_{iyz} \\
\mathcal{J}_{ixz} & \mathcal{J}_{iyz} & \Delta \mathcal{J}_{i\parallel}
    \end{pmatrix}
    \label{eq.Jsym}
\end{equation}
and
\begin{equation}
    \vect{\mathcal{J}^a_i}
    =
\begin{pmatrix}
        0 & -D_{iz} & D_{iy} \\
        D_{iz} & 0 & -D_{ix} \\
        -D_{iy} & D_{ix} & 0 \\
\end{pmatrix}
\end{equation}
This allows Eq.~\ref{eq.anisosimp} to be rewritten as
\begin{eqnarray}
    -\sum_i \sum_{j>i} \vect{S_i} \vect{J_{ij}} \vect{S_j} &=&
     -\sum_i \sum_{j>i} J_{ij}\vect{S_i}\cdot \vect{S_j} \nonumber \\
     &&-\left(\sum_i \vect{S_i}\vect{\mathcal{J}^s_i}\vect{M}
     -\frac{1}{2} \sum_i\vect{M}\vect{\mathcal{J}^s_i}\vect{M} 
     \right)\nonumber \\
     && + \sum_i \mathbf{D_i }\cdot (\vect{S_i\times\vect{M})} \nonumber \\
     &&-\frac{1}{2} \sum_i \sum_{j\neq i}
     \vect{\delta S_i}\vect{\mathcal{J}_{ij}}\vect{\delta S_j}
     \label{eq.anisoD}
\end{eqnarray}
where the components of $\vect{\mathcal{J}^a_i}$ have
been arranged as a site-dependent vector, 
$\mathbf{D_i} = (D_{ix},D_{iy},D_{iz})$.
The notation has been chosen to draw parallels with the
Dzyaloshinsky-Moriya interaction (DMI)~\cite{Dzyaloshinsky1958,Moriya1960}, although it is
important to note that the coupling is between $\vect{S_i}$
and the global vector $\vect{M}$, not explicit pairwise interactions.

Suppose now we are in a ferromagnetic state, i.e.\ $\vect{S_i} = \vect{M}$,
$\vect{\delta S_i} = 0$.
Equation~\ref{eq.anisoD} reduces to 
\begin{equation*}
    -\sum_i \sum_{j>i} \vect{S_i} \vect{J_{ij}} \vect{S_j} = -\sum_i \sum_{j>i} J_{ij}M^2 
    -\frac{1}{2} \sum_i \vect{M}\vect{\mathcal{J}^s_i}\vect{M}
\end{equation*}
i.e.\ an isotropic term and an anisotropic term involving the symmetric
part of  $\vect{\mathcal{J}_i}$.
Although at this point we could carry out the sum
over $\vect{\mathcal{J}^s_i}$, we choose not to do this
and instead write out the atom-resolved contributions
$\vect{M}\vect{\mathcal{J}^s_i}\vect{M}/2$.
Equation~\ref{eq.Jsym} allows us
to re-express these in terms of 
the global magnetization angles $\theta$ and $\phi$, giving
the anisotropic contribution from site $i$ as
\begin{eqnarray}
-\frac{1}{2} \left[\Delta \mathcal{J}_{i\parallel} \left(\frac{3}{2} \cos^2\theta - \frac{1}{2}\right) +\mathcal{J}_{ixz} \sin2\theta \cos\phi
 \nonumber \right.\\
\left.
+\mathcal{J}_{iyz} \sin2\theta \sin\phi
+ \Delta \mathcal{J}_{i\perp} \sin^2\theta \cos2\phi
+\mathcal{J}_{ixy} \sin^2\theta \sin2\phi
  \right]\nonumber 
\end{eqnarray}
Up to an additive constant, this expression is identical to the expression
assumed for the single-ion anisotropy based on the $l=2$ spherical harmonics,
Eq.~\ref{eq.angle2}, with the coefficients mapped to each other through
relations such as $\mathcal{A}_i = -3\Delta\mathcal{J}_{i\parallel}/4$ (the
full set of relations is given in the first row of Table~\ref{tab.expressions}).

We now make the key observation that since $\vect{\mathcal{J}_i}$ is obtained as 
a sum over all neighbouring sites and describes the coupling to a global spin
vector, its symmetry properties can only be determined by those of site $i$.
This means that we can apply the same considerations 
as for Model 1: for various
symmetry operations $\mathcal{O}$, the transformed tensor
$\mathcal{O}\vect{\mathcal{J}_i}\mathcal{O}^T$ will either be invariant
or be equal to another tensor $\vect{\mathcal{J}_j}$ where $j$ is in the same sublattice.
Of course, the equivalence of the coefficients mean that the same relations
derived for Model 1 and given in Table~\ref{tab.expressions} apply also to
the $\vect{\mathcal{J}_i}$ elements of Model 2.

The symmetry analysis also constrains the $\mathbf{D_i}$ vectors, in some
cases forcing them to be zero ($4e$), to align along the $c$-axis ($4c$) or
to lie in the $ab$-plane ($8j$).
These vectors have no effect on the energy of a ferromagnetic state, since $\vect{S_i}\times\vect{M} = 0$.
However, as we show in the next section, they do generate torques
on the individual moments.

\subsection{Torques for a ferromagnetic state}

The first-principles torque method provides access to the quantities $\partial E/\partial \theta _i$
and $\partial E/\partial \phi_i$, where $E$ is the total energy.
Assuming a ferromagnetic reference state, we obtain
\begin{eqnarray}
\left.\frac{\partial E}{\partial\theta_i}\right|_{\vect{S_i}=\vect{M}} &=& -\mathcal{A}_i \sin 2\theta
                 +2\mathcal{B}_i \cos2\theta \cos\phi \nonumber \\
&&                 +2\mathcal{C}_i \cos2\theta \sin\phi
                   +\mathcal{E}_i \sin2\theta \cos2\phi\nonumber \\
&&                 +\mathcal{F}_i \sin2\theta \sin2\phi\nonumber \\
\left.\frac{\partial E}{\partial\phi_i}\right|_{\vect{S_i}=\vect{M}} &=& 
                 -\mathcal{B}_i \sin2\theta \sin\phi
                 +\mathcal{C}_i \sin2\theta \cos\phi\nonumber \\
&&                 -2\mathcal{E}_i \sin^2\theta \sin2\phi
                 +2\mathcal{F}_i \sin^2\theta \cos2\phi\nonumber 
\label{eq.torqssinglesite}
\end{eqnarray}
for Model 1, and 
\begin{eqnarray}
    \left.\frac{\partial E}{\partial \theta_i}\right|_{\vect{S_i}=\vect{M}}
    &=&
    \frac{3}{4} \Delta\mathcal{J}_{i\parallel}\sin2\theta - \mathcal{J}_{ixz}
    \cos2\theta\cos\phi \nonumber \\
    &&-\mathcal{J}_{iyz}\cos2\theta\sin\phi 
    -\frac{1}{2} \Delta\mathcal{J}_{i\perp}\sin2\theta\cos2\phi \nonumber \\
    &&-\frac{1}{2}\mathcal{J}_{ixy}\sin2\theta\sin2\phi  \nonumber \\
    &&+ D_{ix}\sin\phi - D_{iy}\cos\phi \nonumber \\
   \left.\frac{\partial E}{\partial\phi_i}\right|_{\vect{S_i}=\vect{M}} &=& 
   \frac{1}{2}\mathcal{J}_{ixz}\sin2\theta\sin\phi-
   \frac{1}{2}\mathcal{J}_{iyz}\sin2\theta\cos\phi \nonumber \\
   && + \Delta \mathcal{J}_{i\perp}\sin^2\theta \cos2\phi
   -\mathcal{J}_{ixy}\sin^2\theta\cos2\phi
   \nonumber \\
   &&+\frac{D_x}{2}\sin2\theta\cos\phi + \frac{D_y}{2}\sin2\theta\sin\phi  \nonumber \\
   &&-D_z\sin^2\theta 
   \label{eq.torqsaniso}
\end{eqnarray}
for Model 2.
The expressions are almost identical, but crucially there are additional contributions in Model
2 due to $\mathbf{D_i}$.
Hence the site-resolved torques should provide a clear
comparison between models.

\subsection{Torque calculations: computational details}

To numerically test the two models, we have carried
out first-principles torque calculations using the 
methodology described in Ref.~\cite{Staunton2006}
and implemented in the \texttt{MARMOT} code~\cite{Patrick2022}.
Five sets of calculations were carried out, all
on the ferromagnetic state with moments aligned
to a common direction $\vect{M}$.
In sets 1--3, 
$\partial E/\partial \theta_i$ was calculated for a series
of $\theta$ values between 0--90$^\circ$, 
whilst the angle $\phi$ was fixed
at values 0$^\circ$, 45$^\circ$ and 90$^\circ$ for the
respective sets.
In sets 4 and 5, $\partial E/\partial \phi_i$ was 
calculated for $\phi$ varying between 0--90$^\circ$,
with $\theta$ fixed either at 90$^\circ$ and 45$^\circ$.

\begin{table}
\begin{tabular}{|c|c|c|c|c|}
\hline
Atom&Wyckoff Label&$x$&$y$&$z$\\
\hline
\multirow{ 6}{*}{Fe}&$16k_1$ & 0.0671 & 0.2765 & 0.1269 \\
&$16k_2$ & 0.0379 & 0.3587 & 0.3237 \\
&$8j_1$ & 0.0979 & 0.0979 & 0.2951 \\
&$8j_2$ & 0.3174 & 0.3174 & 0.2535 \\
&$4c$ & 0.0 & 0.5 & 0.0 \\
&$4e$ & 0.0 & 0.0 & 0.1144 \\
\hline
\multirow{ 2}{*}{$R$}&4g&0.2313&0.7687&0.0\\
&$4f$&0.3585&0.3585&0.0 \\
\hline
B & $4f$ & 0.1243 & 0.1243 & 0.0 \\
\hline
\end{tabular}
\caption{Atomic coordinates\label{tab.coords}, as determined
experimentally for Nd$_2$Fe$_{14}$B at 300~K~\cite{Isnard1995}.}
\end{table}

The calculations were carried out for 
Y\textsubscript{2}Fe\textsubscript{14}B using lattice
parameters of $a = 8.810$~\AA, $c = 12.203$~\AA,
and the internal parameters listed in Table~\ref{tab.coords}.
These structural parameters were measured experimentally
for 
Nd\textsubscript{2}Fe\textsubscript{14}B~\cite{Huang1987,Isnard1995}
and used in previous work modeling the Fe anisotropy
in that material~\cite{Bouaziz2023}.
The atom-centered potentials were obtained using the
Korringa-Kohn-Rostoker (KKR) formulation of DFT, using
the local-density and 
atomic sphere approximations (ASA) and a reciprocal
space sampling of $6\times6\times4$, as implemented
in the \texttt{HUTSEPOT} code~\cite{Vosko1980,Daene2009}.
The reciprocal space sampling for the torque calculation
was much higher due to the adaptive algorithm implemented
in \texttt{MARMOT}~\cite{Bruno1997,Patrick2022}.

\section{Results}
\label{sec.results}

\subsection{The $4e$ sublattice}

\begin{figure}
\includegraphics[width=0.5\columnwidth]{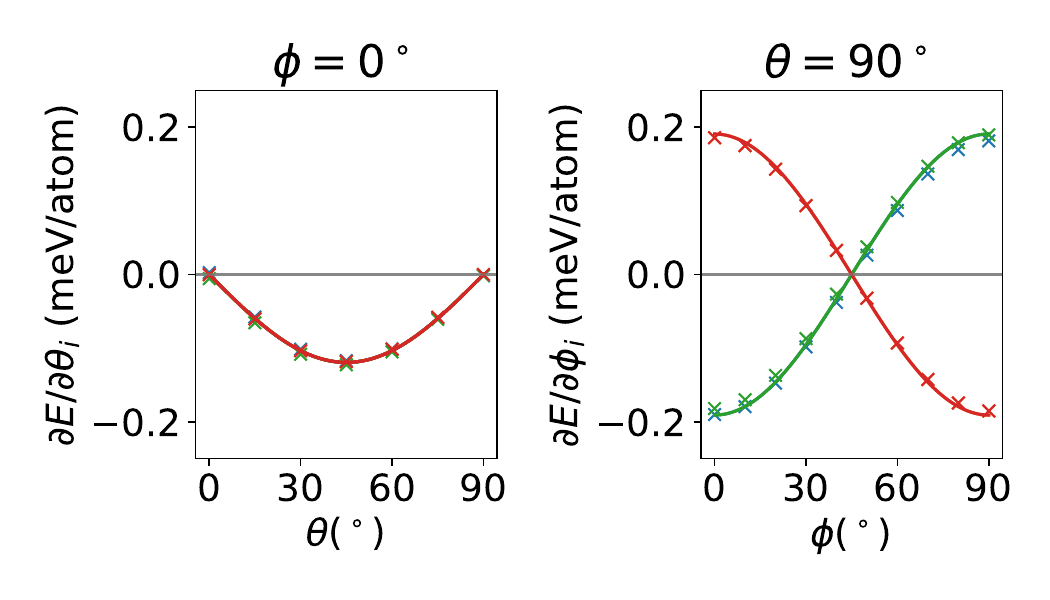}
\caption{DFT-calculated torques (crosses) and Model 1 fits (solid lines)
for Fe atoms in the $4e$ sublattice, for calculation sets 1 and
4 (left and right). \label{fig.4e}}
\end{figure}

We begin with the simplest $4e$ sublattice.
Due to its zero $\mathbf{D}$-vector (Table~\ref{tab.expressions})
Models 1 and 2 predict identical angular dependence
of the torque.
Furthermore, there are only two nonzero coefficients,
$\kappa_{20}$ and $\kappa_{22}$.
For calculation set 1 ($\partial E/\partial\theta_i$ for
$\phi$ fixed at 0$^\circ$) $\kappa_{22}$ plays no role,
and all four atoms are predicted to have identical torques
of $-\kappa_{20} \sin^2\theta$.
As shown in Fig.~\ref{fig.4e} this is exactly what is
found in the calculations, with the data fitted very well
with $\kappa_{20}$ = 0.119~meV/atom.
The figure also shows the results of calculation set 4 
($\partial E/\partial\phi_i$ for
$\theta$ fixed at 90$^\circ$).
In this case the torque is predicted to split the
atoms into two groups, 
following $\pm2\kappa_{22}\cos2\phi$.
Again, we see excellent agreement between the numerical
calculation and the analytical expression, with
$\kappa_{22}$ = -0.095~meV/atom.

\subsection{The $4c$ sublattice}

\begin{figure}
\includegraphics[width=0.5\columnwidth]{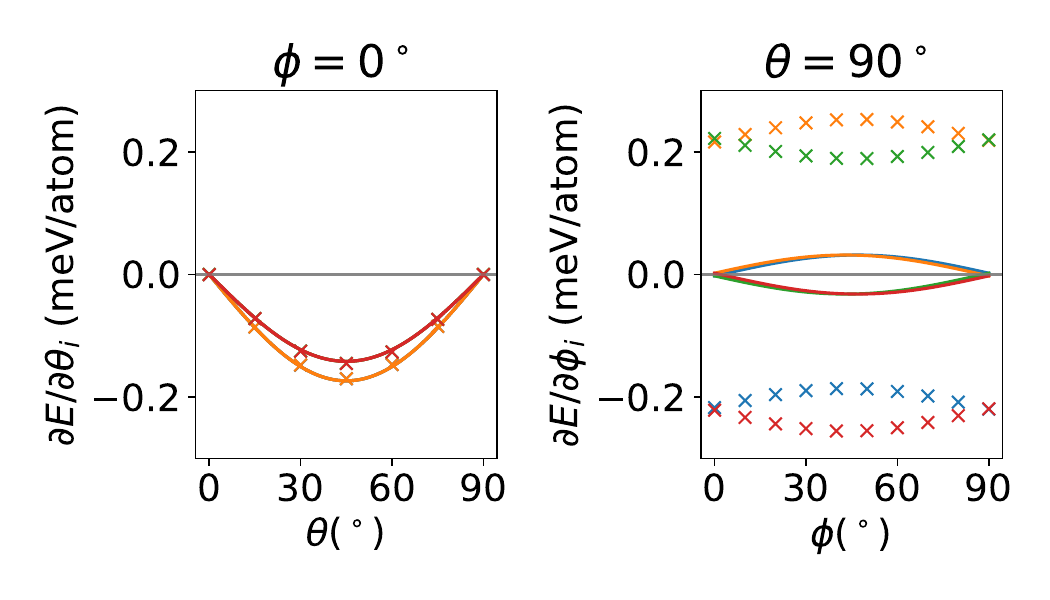}
\caption{The equivalent plots to Fig.~\ref{fig.4e} for Fe
atoms in the $4c$ sublattice.
The DFT-calculated torques in calculation set 4 (right
panel) evidently contain a $\phi$-independent contribution
which is not accounted for in Model 1.
\label{fig.4c}}
\end{figure}

Figure~\ref{fig.4c} shows the results of the same
calculation sets for the $4c$ sublattice, compared
to the analytical expressions predicted by Model 1.
The fits for set 1 are excellent, correctly capturing the
split torques due to the $\kappa_{22c}$ term
(Eq.~\ref{eq.fit_alltorques}).
However, there is a large discrepancy between the calculations
and the analytical fits for set 4.
The calculations find that each atom in the $4c$ group has
its own torque curve, whilst the Model 1 analytical expression
predicts two groups sharing the same torques.
Although the shape of the curves is reasonable, there is clearly
a $\phi$-independent additional contribution present in the 
calculations.

The observed behavior can be explained using Model 2, noting
that the symmetry of 
$4c$ sublattice permits a nonzero $\mathbf{D_i}$
vector pointing along the $z$ axis (Table~\ref{tab.expressions}).
According to Eq.~\ref{eq.torqsaniso} this $\mathbf{D_i}$
vector will have no effect on $\partial E/\partial \theta_i$
torques but will introduce an additional $-D_z\sin^2\theta$
term to $\partial E/\partial \phi_i$ torques.
For set 4, $\theta$  is fixed at $90^\circ$ 
so we expect an additional constant
contribution to the torques, positive or negative depending
on the atom's $\mathbf{D_i}$ vector in Table~\ref{tab.expressions},
exactly as observed.

\subsection{Model 2 for all sublattices}

\begin{figure*}
\includegraphics[width=\columnwidth]{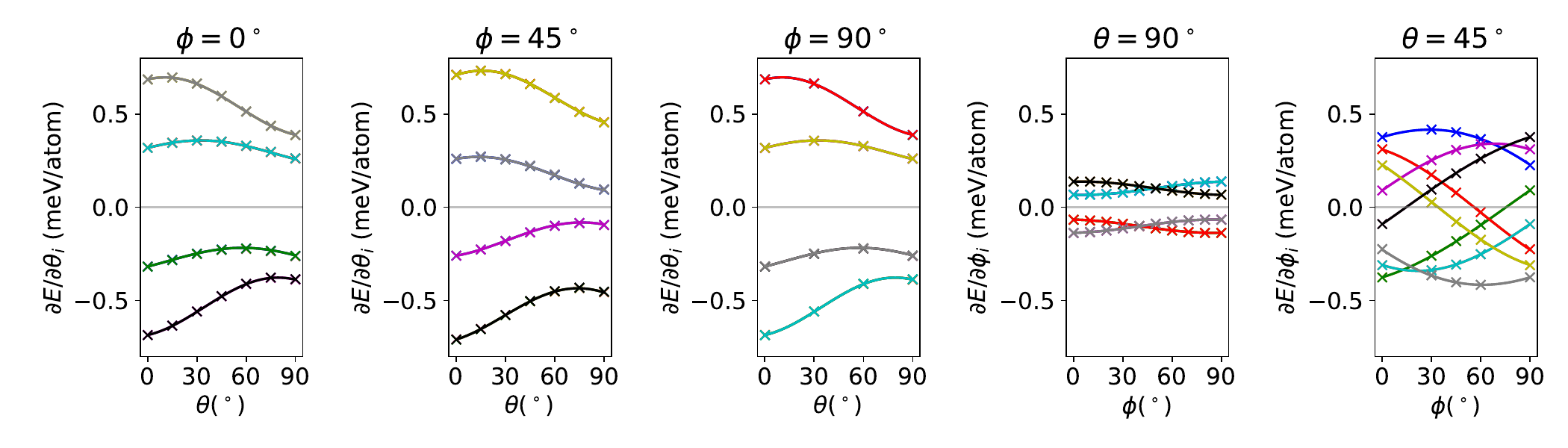}
\caption{The comparison of DFT-calculated torques (crosses) and
Model 2 fits (solid lines) for Fe atoms in the $16k_2$ sublattice,
for all five calculation sets.
Set 5 (rightmost plot) shows the 16 atoms split into 8 distinct
curves, with each atom degenerate with its inversion partner.
\label{fig.16k2}}
\end{figure*}

Considering all sublattices and all calculation sets,
we find that the additional terms in Model 2 deriving from 
anisotropic exchange are essential to fitting 
the first-principles torque
calculations.
Figure~\ref{fig.16k2} demonstrates the model's excellent 
performance
with the $16k_2$ sublattice across the five calculation sets.
We highlight set 5
($\partial E/\partial\phi_i$ for
$\theta$ fixed at 45$^\circ$).
as a particularly interesting example
where the $\mathbf{D_i}$-vector is responsible for a 
$\phi$-dependent contribution.
The 16 atoms are split into 8 groups consisting of an atom
and its inversion partner.
As required by the tetragonal symmetry and the quadratic ($l=2$)
nature of the expansion, these eight distinct curves sum to zero.

\begin{table*}
\begin{tabular}{|c|c|c|c|c|c|c|c|c|c|c|c|}
\hline
& $\kappa_{20}$  & $\kappa_{21}$ & $\kappa_{21c}$ &$\kappa_{21s}$ &
$\kappa_{22}$ & $\kappa_{22c}$ &$\kappa_{22s}$ &
$D_{xy}$ & $D_x$ & $D_y$ & $D_z$ \\
\hline
$16k_1$ & -0.046&---&-0.002&0.013&---&-0.015&-0.009&---&-0.799&-0.432&0.064\\
$16k_2$ & -0.061&---&-0.015&0.074&---&0.003&-0.018&---&0.537&0.288&-0.102\\
$8j_1$ & -0.010&0.037&---&---&-0.052&---&---&-0.099&---&---&--- \\
$8j_2$ & -0.032&0.044&---&---&-0.047&---&---&0.686&---&---&--- \\
$4c$ & 0.157&---&---&---&---&-0.016&-0.001&---&---&---&0.220\\
$4e$ & 0.119&---&---&---&-0.095&---&---&---&---&---&---\\
\hline
\end{tabular}
\caption{Values of anisotropy constants determined from
fitting of first-principles torque calculations.
All quantities are in units of meV/atom. \label{tab.values}
}
\end{table*}

The torque data and fits for all sublattices are provided as 
Supplemental Material~\cite{suppinfo}).
The fitted values of the anisotropy constants 
are provided in Table~\ref{tab.values}.
The total anisotropy energy $E^{[100]} - E^{[001]}$
is equal to $-\sum \kappa_{20} = $0.24~meV/FU, as found
previously using the same methodology~\cite{Bouaziz2023}.

\section{Discussion}
\label{sec.discussion}

\subsection{Relevance to previous work}

We recall the original motivation for our study was
to explain the splittings in site-resolved anisotropy
energies observed
in Refs.~\cite{Miura2014,Torbatian2014} between atoms
in certain sublattices for the energy difference
$E^{[100]} - E^{[001]}$.
Both Model 1 and Model 2 explain these
splittings.
For  $E^{[100]} - E^{[001]}$, Eq.~\ref{eq.angle2} 
gives the contribution per atom
as
$\mathcal{E}_i - \mathcal{A}_i$.
Table~\ref{tab.expressions} shows that only
the $16k$ and $4c$ sublattices have nonzero 
$\mathcal{E}_i$ values, and that within these sublattices,
these are equally split between positive and negative
values.
Hence half of the atoms in these sublattices will
contribute $\kappa_{22c}-\kappa_{20}$ to the anisotropy
energy, and the other half contribute $-\kappa_{22c}-\kappa_{20}$.
Averaging the contributions over all atoms in the sublattice
will retrieve the single value $-\kappa_{20}$, which is equal to $D_i$.
However, since Fig.~1 of
Ref.~\cite{Miura2014} does not show both of 
the split contributions,
the averaging cannot be carried out and the appropriate 
$D_i$ values cannot be obtained.
The factor-of-seven discrepancy identified in Sec.~\ref{sec.theory}
arises because $\kappa_{22c}-\kappa_{20}$ 
(or $-\kappa_{22c}-\kappa_{20}$)
is being used
as $D_i$, rather than $-\kappa_{20}$.

\subsection{Recommendations for future ASD calculations}

A key question which arises from our results is, how
should Fe anisotropy be treated in future ASD
studies of Nd$_2$Fe$_{14}$B?
We propose three options:
\subsubsection{Original ASD Hamiltonian, averaged $D_i$ values}
The simplest approach would be to keep the same Hamiltonian
and anisotropy expression, Eqs.~\ref{eq.Di}~and~\ref{eq.HASD},
but use $D_i$ values which take into account the splitting
of the $16k$ and $4c$ sublattices.
This approach requires averaging
over the split values.
This can be done directly 
for the results of Ref.~\cite{Torbatian2014}, whilst for
Ref.~\cite{Miura2014} it would be necessary to obtain 
from the authors the 
extra data which does not appear in the 
published paper.
With this approach, using Ref.~\cite{Torbatian2014} gives
updated $D_i$ values (for 
$16k_1$, $16k_2$,
$8j_1$, $8j_2$, $4c$ and $4e$) as
0.05, 0.13, 0.23, 0.06, -0.40, and 0.07~meV/atom, giving
a total MAE of $\sim$1~meV/FU, similar to
that measured experimentally for Y$_2$Fe$_{14}$B close to
zero temperature~\cite{Hirosawa1986,Bouaziz2023}.
Such a change should be easy to make and would quickly 
evaluate to what extent the treatment of the Fe anisotropy
changes the qualitative or quantitative conclusions of the
previous ASD studies.

\subsubsection{Direct implementation of Model 1}

A second approach, which is not expected to 
require much effort to implement, is to retain
the ASD Hamiltonian of Eq.~\ref{eq.HASD} but replace
the Fe site anisotropy expression (Eq.~\ref{eq.Di})
with the more complicated single-ion expressions derived
in Model 1 (Eq.~\ref{eq.fit_alltorques}).
In this approach, the anisotropy of each site would
truly reflect its symmetry.
A difficulty of this approach is that the 
anisotropy constants presented in Table~\ref{tab.values}
only sum up to a value of 0.24~meV/FU, 4--5 times smaller
than the experimental value for 
Y$_2$Fe$_{14}$B~\cite{Hirosawa1986}.
We attribute this difference is to our use of the ASA,
noting that full-potential KKR calculations give larger
anisotropy~\cite{Bouaziz2023}.
A solution could be to scale the values in Table~\ref{tab.values}
to reproduce the experimental anisotropy.
As for the previous approach, it would be interesting to 
test whether this different description of the Fe anisotropy
has an observable effect on the properties calculated with
ASD, particularly domain wall structures.

\subsubsection{Mean-field implementation of Model 2}
 
 The most ambitious approach would be to replace
 the isotropic Fe-Fe exchange $J_{ij}\vect{S_i}\cdot\vect{S_j}$
 and single-ion Fe anisotropy $E_{s,i}(\vect{S_i})$
 in Eq.~\ref{eq.HASD}
 with the anisotropic exchange term $\vect{S_i}\vect{J_{ij}}\vect{S_j}$
 given in Eq.~\ref{eq.anisoD}.
 However, our strategy of determining anisotropy coefficients 
 through torque calculations
 on the ferromagnetic reference state does not allow us to 
 determine the coefficients $\vect{\mathcal{J}_{ij}}$ 
 which appear in the final term of Eq.~\ref{eq.anisoD}.
 As noted in Sec.~\ref{sec.model2} it would be a formidable
 task to determine these, and an ASD model which included them
 would be difficult to interpret physically due to its inherent complexity.

 As an alternative, we propose an approach
 which is simpler to implement and interpret: we neglect
 the final term in Eq.~\ref{eq.anisoD} involving the product of fluctuations,
 and take as the reference
 vector $\vect{M}$ the ferromagnetic order parameter for
 the Fe atoms, which could be evaluated at each timestep as an average
 over a suitably large supercell,
 \begin{equation}
     \vect{M} = \frac{1}{N_{\mathrm{Fe}}}\sum_{i\in\mathrm{Fe}} \vect{S_i}.
     \label{eq.Mav}
 \end{equation}
Having made this approximation, the new ASD Hamiltonian has the following form:
\begin{eqnarray}
    \mathcal{H} &=&
     -\sum_i \sum_{j>i} J_{ij}\vect{S_i}\cdot \vect{S_j} \nonumber \\
     &&-\left(\sum_{i\in\mathrm{Fe}} \vect{S_i}\vect{\mathcal{J}^s_i}\vect{M}
     -\frac{1}{2} \sum_{i\in\mathrm{Fe}}\vect{M}\vect{\mathcal{J}^s_i}\vect{M} 
     \right)\nonumber \\
     && + \sum_{i\in\mathrm{Fe}} \mathbf{D_i }\cdot (\vect{S_i\times\vect{M})} \nonumber \\
     &&
    + \sum_{i\in R} E_{Rs,i}(\vect{S_i}) - \mu_0\sum_i{m_i}\vect{h}\cdot\vect{S_i}
    \label{eq.HASDnew}
\end{eqnarray}
The terms in Eq.~\ref{eq.HASDnew} correspond to isotropic exchange between
all atoms,
anisotropic symmetric and antisymmetric exchange between
Fe atoms, single-ion anisotropy for $R$ atoms, and the Zeeman interaction for
all atoms.
We note that our mean-field approximation has only been applied to the anisotropic 
exchange terms; the dominant, isotropic exchange is treated at the same level as
in the original ASD Hamiltonian, Eq.~\ref{eq.HASD}.

We believe Eq.~\ref{eq.HASDnew} represents the best balance between computational
tractability and capturing the itinerant electron anisotropy of the Fe moments,
including the previously uninvestigated $\mathbf{D_i }$ contributions.
These will drive non-collinearity within the sublattices, although we expect the effect
to be small due to the strong isotropic exchange.
The coefficients which appear in the new terms of Eq.~\ref{eq.HASDnew}
have all been calculated in this work, presented in Tables~\ref{tab.expressions}
and~\ref{tab.values}.
As noted for Model 1, due to our use of the ASA 
it would be necessary to scale these factors to reproduce the experimental anisotropy
of Y$_2$Fe$_{14}$B.
A more satisfactory solution would be to 
carry out full-potential torque calculations, but this is beyond the scope of our current study.

Finally, we note that one could in principle retain the single-ion description of Fe 
anisotropy from Model 1, but add to it the anisotropic antisymmetric exchange contribution
from $\mathbf{D_i }$ (using a suitable approximation for $\vect{M}$, Eq.~\ref{eq.Mav}).
One could also suppose the Fe anisotropy to have some single-ion 
character, since our use of the ferromagnetic reference state does not allow us to 
distinguish between this and (symmetric) anisotropic exchange.
However, what our calculations do confirm is that the antisymmetric and symmetric
contributions to the torque are of similar magnitudes, and the former cannot be
explained by the single-ion model.
Therefore, we do not believe there is strong justification for 
including the antisymmetric contribution $\mathbf{D_i}$ without the symmetric
contribution $\vect{\mathcal{J}^s_i}$.

\subsection{Temperature-dependent anisotropy for ferromagnets}

We conclude our discussion by pointing out an interesting
consequence of the proposed ASD Hamiltonian, Eq.~\ref{eq.HASDnew},
when applied to a simple itinerant electron ferromagnet, with only
one magnetic sublattice.
The archetypal example of such a system is $L1_0$-ordered 
FePt~\cite{Staunton2004}.
Applying the analysis described in this study, 
the high symmetry of the Fe site leads to the
$\mathbf{D_i}$ vector and all coefficients except
$\kappa_{20}$ equalling zero.
Furthermore, since there is only one magnetic sublattice, the order
parameter $\vect{M}$ is equal to the thermal
average of $\vect{S_i}$, i.e.\ $\langle\vect{S_i}\rangle_T$.
This means that if one considers the thermal average
of Eq.~\ref{eq.HASDnew}, one obtains
\begin{equation}
    \langle \mathcal{H}\rangle_T = \mathrm{isotropic \ term} 
    + M^2  \times K_0\sin^2\theta
    \label{eq.KFePt}
\end{equation}
where $K_0$ is the conventional anisotropy constant at zero temperature $(M=1)$.
Therefore, Eqs.~\ref{eq.HASDnew} and~\ref{eq.KFePt} produce the well-known
$M^2$ dependence of the anisotropy energy, observed experimentally for
FePt~\cite{Okamoto2002,Thiele2002}.

Whilst this is a satisfying result for FePt and similar systems,
Eq.~\ref{eq.HASDnew} does not explain the
temperature dependence of anisotropy in ferromagnets which does not
follow $M^2$ behavior, such as in MnBi~\cite{Patrick20222} 
or Co-Cr-X alloys~\cite{Inaba2000}.
However, the perturbative analysis of Ref.~\cite{Miura2022} 
shows how the $M^2$ behavior
arises in the fully-itinerant (wide band) limit.
For materials where the electrons have a more localized character,
a single-ion type term may need to be included as well, which would
lead to a more complicated temperature dependence.

\section{Conclusion}
\label{sec.conclusions}

We have derived and parameterized two models of the contribution from
the Fe atoms
to the magnetocrystalline anisotropy energy of $R_2$Fe$_{14}$B
permanent magnets.
Model 1 is an extension of the single-ion model used
in previous works, using expressions which fully reflect
the symmetry of each Fe site.
Model 2 assumes an anisotropic exchange interaction, written
in such a way which produces site-resolved contributions in
the ferromagnetic reference state.
A key prediction of Model 2 is an antisymmetric 
exchange contribution, akin to a DMI term, but with the coupling
between a single spin and a global vector $\vect{M}$ rather
than between a pair of spins.
First-principles torque calculations reveal that this antisymmetric
contribution is indeed present, and only Model 2 fits the data across
all sublattices and datasets.
We have shown how our expressions are compatible with previous 
DFT calculations of site-resolved anisotropy energies,
and we have also proposed practical strategies to carry out future ASD simulations
on $R_2$Fe$_{14}$B based on our findings.
As indicated by our discussion of FePt, we believe this methodology 
has further applications outside $R_2$Fe$_{14}$B.
In particular, our method provides a route to carrying out ASD simulations of
rare-earth-lean or rare-earth-free permanent magnets based on itinerant
magnetism, particularly where the complexity of the crystal structure
prevents a full determination of all  $\vect{\mathcal{J}_{ij}}$ tensors.

\begin{acknowledgments}
We thank Y.\ Miura, J.\ B.\ Staunton
and J.\ Bouaziz for useful discussions.  Partial funding for this work
was provided by The Armourers \& Brasiers' Company (UK).
\end{acknowledgments}

\bibliography{papers}

\end{document}